\documentstyle[psfig,epsfig,aps]{revtex}

\def\beq{\begin{equation}}
\def\eeq{\end{equation}}
\def\lsim{\raise0.3ex\hbox{$\;<$\kern-0.75em\raise-1.1ex\hbox{$\sim\;$}}}
\def\gsim{\raise0.3ex\hbox{$\;>$\kern-0.75em\raise-1.1ex\hbox{$\sim\;$}}}

\def\apj#1#2#3{ Astrophys.\ J.\ {\bf #1}, #3 (#2)}
\def\ib#1#2#3{ ibid.\ {\bf #1}, #3 (#2)}
\def\mn#1#2#3{ Mon.\ Not.\ R.\ Astron.\ Soc.\ {\bf #1}, #3 (#2)}
 
\def\np#1#2#3{ Nucl.\ Phys.\ {\bf #1}, #3 (#2)}

\def\pl#1#2#3{ Phys.\ Lett.\ {\bf #1}, #3 (#2)}
\def\prd#1#2#3{ Phys.\ Rev.\ D {\bf #1}, #3 (#2)}

\def\prl#1#2#3{ Phys.\ Rev.\ Lett.\ {\bf #1}, #3 (#2)}

\def\rpp#1#2#3{ Rept.\ Prog.\ Phys.\ {\bf #1}, #3 (#2)}

\begin{document}

\title{Cosmological measurement of neutrino mass in the presence of
leptonic asymmetry}

\author{Julien Lesgourgues$^1$, Sergio Pastor$^1$ and Simon Prunet$^2$}
\address{
$^1$ SISSA--ISAS and INFN, Sezione di Trieste,
Via Beirut 4, 34014 Trieste, Italy}
\address{
$^2$ CITA, 60 St George Street, Toronto, Ontario M5S 3H8, Canada
}

\maketitle

\vskip0.7cm

\begin{abstract}
We show that even the smallest neutrino mass consistent with the
Super--Kamiokande data is relevant for cosmological models of
structure formation and cosmic microwave background (CMB)
anisotropies, provided that a relic neutrino asymmetry exists. We
calculate the precision with which a $0.07$ eV neutrino mass could be
extracted from CMB anisotropy and large-scale structure data by the
future Planck satellite and Sloan Digital Sky Survey. We find that
such a mass can be detected, assuming a large relic neutrino asymmetry
still allowed by current experimental data. This measurement of the
absolute value of the neutrino mass would be crucial for our
understanding of neutrino models.
\end{abstract}

\section{Introduction}

One of the most intriguing questions in cosmology is the possibility
of having an asymmetry in the number of leptons and antileptons in the
Universe. This asymmetry is restricted to be in the form of neutrinos
from the requirement of universal electric neutrality.  A large
neutrino asymmetry is not excluded by current observational data of
primordial abundances of light elements, cosmic microwave background
(CMB) anisotropies and large scale structure in the Universe.

If a relic asymmetry exists, the corresponding neutrinos, called
degenerate, are characterized by the dimensionless degeneracy
parameter $\xi \equiv \mu / T_{\nu}$, where $\mu$ is their
chemical potential and $T_{\nu}$ their temperature.  The energy density
of degenerate neutrinos is much larger than the one of standard
neutrinos, and is a function of $\xi$.

{}From a theoretical point of view, in most particle physics models
the leptonic asymmetry is naturally of the same order as the baryonic
one, i.e.~one part in $10^{9}-10^{10}$ as required by big bang
nucleosynthesis (BBN) \cite{Sarkar96}. However, there are some
specific scenarios where the leptonic asymmetry can grow up to large
values in the early universe \cite{Casas}, while the baryonic one
remains small.  Some examples include lepton asymmetries created by an
Affleck-Dine mechanism \cite{AF} or by active-sterile neutrino
oscillations \cite{Foot}, which allow the neutrino asymmetry to reach
order one before neutrinos decouple from the rest of the plasma.
In general, these generating mechanisms create a different asymmetry
for different neutrino flavors.

{}From an observational point of view, BBN constrains the neutrino
degeneracy to be at most of order one for the electronic neutrino
($-0.06 \lsim \xi_{\nu_e} \lsim 1.1$) \cite{Kang}, but is compatible
with larger degeneracies for $\nu_{\mu}$ or
$\nu_{\tau}$. Interestingly, the current CMB anisotropy data are
compatible with a large neutrino asymmetry $\xi \sim 3.5$ in the
framework of the standard cold dark matter (CDM) cosmological
scenario, as shown in \cite{Sarkar}. More recently, two of us have
shown in a systematic analysis \cite{paper1} that this conclusion also
applies to flat models with a cosmological constant ($\Lambda$CDM),
even when CMB data are combined with constraints on the matter power
spectrum (for an earlier discussion see \cite{Larsen}).  This analysis
is based on the data available when \cite{paper1} was submitted. We
have checked that more recent data such as TOCO \cite{TOCO} and
BOOMERANG 97 \cite{Boomerang} are still compatible with our previous
upper bound, $\xi \lsim 3.5$.

In the case of massless degenerate neutrinos, the only relevant effect
of $\xi$ is to increase the total density of radiation, and to
postpone the time of equality between radiation and matter. This
modification has got large observable effects: it boosts the first CMB
peak amplitude, shifts all peaks to smaller scales, and suppresses
matter fluctuations on small scales \cite{paper1}. However, it can be
simply described by introducing an effective number of massless
neutrino families
\begin{equation}
N (\xi) \equiv 3 + \frac{30}{7} \left (\frac{\xi}{\pi}\right )^2 +
\frac{15}{7} \left (\frac{\xi}{\pi}\right )^4
\label{Nxi}
\end{equation}
which is as large as $5$ for $\xi \simeq 2$. This excess in the
effective number of neutrinos would wash out the small corrections
that arise in the standard model (slight heating of neutrinos by
$e^+-e^-$ annihilations and finite-temperature QED effects), whose
effects on the CMB were considered in \cite{Lopez}.

The analysis in \cite{paper1} also included the case of massive
degenerate neutrinos, to which we adapted the Boltzmann code {\sc
cmbfast} by Seljak and Zaldarriaga \cite{SelZal}, that calculates the
radiation and matter power spectra.  It appeared that combining the
asymmetry with a small mass for one family of neutrinos had some
subtle effects, that can not be parametrized simply with $N(\xi)$.
For instance, the suppression of small scale matter fluctuations
caused by the free-streaming of neutrinos (when they become
non-relativistic) is more efficient in presence of an asymmetry, due
to the enhanced average momentum of the degenerate neutrino. Also, by
combining a mass and a degeneracy for the same neutrino family, we
reach a bigger neutrino density today than by introducing these
parameters separately, or for different families.  This point has
very interesting phenomenological consequences.

The evidence for neutrino oscillations from Super--Kamiokande
\cite{SK}, if explained by standard $\nu_{\mu} \leftrightarrow
\nu_{\tau}$ oscillations (for recent reviews, see for instance
\cite{nurevs}), requires differences of squared masses of the order
$\Delta m^2 \simeq (1-8)\times 10^{-3}$ eV$^2$. This determines a
lower limit on the value of the neutrino mass, $m \geq 0.03-0.09$
eV. This bound is saturated in the case when there is a hierarchy in
the neutrino mass pattern, i.e.~when the two neutrino masses are very
different. In the present work we consider as a typical value
$m = m_{SK}= 0.07$ eV.

However such very light neutrinos only make a small
contribution to the present energy density of the Universe, of the
order $\Omega_{\nu}=0.00075 h^{-2}$ for a dimensionless Hubble
parameter $h=H/(100~\mbox{km s}^{-1}\mbox{Mpc}^{-1})$, while at the
same time they have no visible effect on the power spectra of matter
and CMB anisotropies. Therefore one concludes that a $0.07$ eV
neutrino is of little relevance for cosmology. But this conclusion is
modified when one considers the combined effects of mass and
degeneracy.  For instance, the present energy density of neutrinos
with $m_{SK}$ and $\xi=3$ is of the same order of magnitude as the one
of baryons \cite{paper1,PalKar}. In such a case the light degenerate
neutrino plays the role of a significant hot dark matter component.

The main motivation of this paper is to address the following
question: are the future CMB experiments sensitive enough to detect a
$0.07$ eV neutrino mass in the presence of a relic neutrino asymmetry?
Such an evidence would be of tremendous importance for our
understanding of neutrino models, since it would probe the absolute
value of the neutrino mass, while neutrino oscillations are sensitive
to the difference of squared masses.  The sensitivity of the future
satellite missions \cite{MAP+PLANCK} Microwave Anisotropy Probe (MAP)
and Planck to heavier neutrinos was considered in \cite{numass}. Since
$\xi$ enhances the effect of the mass, but, on the other hand,
introduces a new degree of freedom in the model, it is not obvious
whether the scheduled experiments have the required sensitivity to
detect a degenerate neutrino mass as small as $m_{SK}$.  From our
calculations we conclude that Planck will be able to detect it,
provided that there exists a large relic neutrino asymmetry, typically
$\xi \gsim 2-3$. This value is close to the one suggested to explain
the production of ultra-high energy cosmic rays beyond the
Greisen-Zatsepin-Kuzmin cutoff \cite{Gelmini}.

In a previous work \cite{Kinney}, Kinney and Riotto already calculated
the precision with which the MAP and Planck satellites could measure a
large degeneracy parameter $\xi \sim {\cal O} (1)$. The effect that we
consider in this work is so tiny that we will skip MAP and focus on
the capabilities of Planck, as well as those of the future Sloan
Digital Sky Survey (SDSS), that will probe the shape of the matter
power spectrum. It is well known that for ordinary massive neutrinos,
combining CMB and Large Scale Structure (LSS) data is crucial for the
mass extraction \cite{numass}; in our case it is even more important,
due to the enhanced free streaming effect on small scale matter
fluctuations.

For completeness, we also consider the case of a slightly heavier
neutrino, with $m=1$ eV. This mass is of the order of magnitude that
could explain the results from the Los Alamos Liquid Scintillation
Neutrino Detector (LSND) \cite{LSND} experiment through neutrino
oscillations, which require $\Delta m^2 \simeq 0.1-1$ eV$^2$. An eV
neutrino mass is also required in cold + hot dark matter models
(CHDM), because it produces $\Omega_{\nu} \gsim 0.01h^{-2}$. It has
been already shown that such a mass could be extracted with $\sim
20$\% precision by Planck + SDSS \cite{numass}. We calculate how much
this result is improved in the presence of a large asymmetry.

\section{The Fisher matrix}

Since the sensitivity of Planck and of the SDSS is already known, it
is possible to assume a ``fiducial'' model, i.e., a cosmological model
that would yield the best fit to the future data, and to forecast the
error with which each parameters would be extracted.  

Starting with a set of parameters $\theta_i$ describing the fiducial
model, one can compute the power spectra of CMB temperature and
polarization anisotropies. Since the anisotropy data consists in
two-dimensional maps of the sky, these power spectra are usually
expanded in multipoles $C_l^X$, where $l$ is the multipole number, and
$X$ is one of the temperature or polarization modes $T,E,TE,B$
\cite{POL}. Simultaneously, one can derive the linear
power spectrum of matter fluctuations $P(k)$, expanded in Fourier
space. Although CMB experiments measure the $C_l^X$'s directly,
redshift surveys such as the SDSS probe the linear power spectrum only
on the largest scales, and modulo a biasing factor $b^2$. For a given
survey, the biasing reflects the discrepancy between the total matter
fluctuations in the Universe, and those actually seen by the
instruments. It is usually assumed to be independent of $k$.

The error $\delta \theta_i$ on each parameter can be calculated from
the reduced Fisher matrix $F_{ij}$, which has two terms.  The first
one accounts for Planck and is computed according to
ref.~\cite{FISH.CMB}, while the second term accounts for the SDSS and
is calculated following Tegmark \cite{Teg}
\begin{eqnarray} 
F_{ij}&=&\sum_{l=2}^{+ \infty} \sum_{X,Y} \frac{\partial C_l^X}{\partial \ln
\theta_i}
{\rm Cov}^{-1} (C_l^X, C_l^Y) \frac{\partial C_l^Y}{\partial \ln \theta_j}
\nonumber \\
&+& 2 \pi \int_{0}^{k_{max}}
\frac{\partial \ln P_{obs} (k)}{\partial \ln \theta_i}
\frac{\partial \ln P_{obs} (k)}{\partial \ln \theta_j} w(k) d \ln k.
\label{fisher.matrix}
\end{eqnarray}
Here ${\rm Cov} (C_l^X, C_l^Y)$ is the covariance matrix of the
estimators of the CMB spectra for Planck, and $w(k)$ is the weight
function for the bright red galaxies sample of the SDSS, taken from
Tegmark \cite{Teg}. We defined $P_{obs}(k) \equiv b^2 P(k)$, and
$k_{max}$ is the maximal wave number on which linear predictions
are reliable.  Following \cite{numass}, we will use either the
conservative value $k_{max} =0.1 h$ Mpc$^{-1}$, or the optimistic but
still reasonable value $k_{max} =0.2 h$ Mpc$^{-1}$.

Inverting $F_{ij}$, one obtains the 1-$\sigma$ error on each parameter,
assuming that all other parameters are unknown
\begin{equation}
\frac{\delta \theta_i}{\theta_i} = (F^{-1})_{ii}^{1/2}.
\end{equation}
It is also useful to compute the eigenvectors of the reduced Fisher
matrix (i.e., the axes of the likelihood ellipsoid in the space of
relative errors). The error on each eigenvector is given by the
inverse square root of the corresponding eigenvalue. The eigenvectors
with large errors indicate directions of parameter degeneracy; those
with the smallest errors are the best constrained combinations of
parameters.

We assume that a best fit to the future Planck and SDSS data (our
``fiducial'' model) is a $\Lambda$CDM model with ten parameters: (1) a
neutrino mass $m=0, 0.07$ or $1$ eV, (2) a neutrino degeneracy $\xi$,
(3) a Hubble parameter $h=0.65$, (4) a baryon density
$\Omega_b=0.015h^{-2}$, (5) a cold dark matter density
$\Omega_{CDM}=0.3$, (6) a primordial spectrum tilt $n=0.98$, (7) a
primordial spectrum normalization, fixed in {\sc cmbfast} by fitting
to COBE, (8) an optical depth to reionization $\tau=0.05$, (9) a
quadrupole tensor-to-scalar ratio $T/S=0.14$, (10) an arbitrary SDSS
bias $b$.  These parameters were chosen in such way that for $m_{SK}$
and $0 \leq \xi \leq 3.5$, the fiducial models pass the observational
tests of \cite{paper1}. These tests are independent of the bias, and
so is the Fisher matrix as can be seen from Eq.~(\ref{fisher.matrix}).

\section{Measuring the degeneracy parameter}

We first consider a fiducial model with $m=0$ and a degeneracy
parameter $\xi$.  Our results are shown in Fig.~\ref{deltaxi}, where
we plot $\delta \xi / \xi$ as a function of $\xi$. For a very large
degeneracy $\xi=3.5$, we find for Planck alone, without polarization,
$\delta \xi / \xi = 2.5 \%$. Such a small error is justified by the
large effect of the degeneracy on the amplitude and shape of the
acoustic peaks \cite{Sarkar,paper1,Kinney}.  It is limited by a small
parameter degeneracy between $\xi$ and $\Omega_{CDM}$. Computing the
eigenvectors of the Fisher matrix, we find that the combination
$\Omega_{CDM}^{0.8}/\xi^{0.6}$ is measured with $2.9 \%$
uncertainty. This is equivalent to a degeneracy between $\xi$ and the
cosmological constant, since when $m=0$, one has that
$\Omega_{CDM}+\Omega_{\Lambda}=1-\Omega_b$.  There is a simple
physical explanation: the only effect of $\xi$ is to change the time
of equality, and this is one of the main effects of
$\Omega_{\Lambda}$. Since this explanation holds not only for the
temperature spectrum, but also for the polarization and matter
spectra, we do not expect to remove this degeneracy by including the
information from polarization and the SDSS: indeed, $\delta \xi / \xi$
is reduced only from $2.5\%$ to $2.3\%$. So, only direct precise
measurements of $\Omega_{CDM}$ and $\Omega_{\Lambda}$ (using
gravitational lensing or supernovae) could improve this already good
result.

Our results are in good agreement with Kinney and Riotto
\cite{Kinney}. Interestingly, they are found to be almost independent of
the mass of the degenerate neutrino. With a significant value of the
mass, one may expect to lose precision on $\xi$, due to possible
degeneracies between $\xi$ and $m$: both parameters boost the
acoustic peaks and suppress power on small scales. However, the two
effects should remain separable, because $\xi$ changes the radiation
density at all times, while $m$ affects it only when the
neutrinos become non-relativistic. Indeed, for $m = 1$ eV, we
check by diagonalizing the Fisher matrix that no degeneracy appears
between $\xi$ and $m$.  By varying $m$ from 0 to $1$ eV, we
find that $\delta \xi/\xi$ increases by only $\sim 10\%$.

\section{Measuring the degenerate neutrino mass}

We introduce one family of massive degenerate neutrinos with
$m_{SK}$. In this case, the results for $\delta m / m$ as a function
of $\xi$ are shown as the upper curves in Fig.~\ref{deltamu}. For
instance, when $\xi=2.75$, the temperature anisotropy measurement by
Planck can bring evidence for $m_{SK}$, but only marginally: $\delta
m/m = 102 \%$. Here the precision is limited by a parameter degeneracy
direction $(T/S)^{0.7} /m^{0.7}$, with eigenvalue $130\%$. Since the
polarization measurement is able to constrain $T/S$ better, when
including it we find $\delta m/m=93\%$. Finally, significant progress
is made with the SDSS, and the final error depends on $k_{max}$, since
the SDSS is sensitive on small scales to the free streaming produced
by degenerate neutrinos.  For $k_{max}=0.1h$ Mpc$^{-1}$ (respectively
$0.2$) we obtain $\delta m / m = 84 \%$ (respectively $59\%$), which
results in a clear detection, especially if we recall that $\partial
C_l/\partial m$ is a strongly non-linear function of $m$ when $m
\rightarrow 0$, so that the above errors, when large, are
overestimated, as pointed out in \cite{numass}. When the neutrino
degeneracy is as large as $\xi=3.5$, we get $\delta m / m= 37 \%$ for
Planck + SDSS.

We also plot in Fig.~\ref{deltamu} the results for $m=1$ eV (lower
curves).  When $\xi \rightarrow 0$, the estimated errors are consistent with
\cite{numass}.  In the presence of a relic neutrino asymmetry, they
are lowered from $\delta m / m=15-20\%$ (for $\xi=0$) to
$3-4\%$ (for $\xi=3$), depending on $k_{max}$.

\section{Conclusions}

We have shown that Planck will be able to detect a neutrino mass of
the order $m_{SK}$, provided that the relic neutrinos are strongly
degenerate, with a degeneracy parameter $\xi \gsim 3$.  The
combination of Planck with the SDSS improves the results and allows a
detection of $m_{SK}$ if $\xi \gsim 2.5$.  Therefore the neutrino mass
suggested by Super--Kamiokande could be relevant for cosmological
models of structure formation and CMB anisotropies.

We have also confirmed that in the massless neutrino case, the
degeneracy parameter $\xi$ can be extracted from the CMB data by
Planck with the precision found by Kinney and Riotto
\cite{Kinney}. For this parameter, we find that the inclusion of the
SDSS data or the addition of a neutrino mass would not significantly
change the results. Last but not least, if the mass of the neutrinos
is of the order of $1$ eV, then even in absence of asymmetry it can be
extracted with Planck and SDSS with the precision found by
\cite{numass}, but the relic neutrino asymmetry allows a more accurate
detection.

The possibility of detecting a neutrino mass and/or a relic neutrino
asymmetry from future cosmological experiments is an example of the
fascinating connection between large scale cosmology and particle
physics.

\section*{Acknowledgments}

J. Lesgourgues and S. Pastor are supported by the European Commission
under the TMR contract ERBFMRXCT960090.

\newpage

\begin{figure}
\caption{$\delta \xi/\xi$ versus $\xi$, for one family of massless
degenerate neutrinos. From top to bottom, the four curves
refer to Planck without and with polarization, and Planck + SDSS with
$k_{max}=0.1$ or $0.2 h$ Mpc$^{-1}$. The results are almost identical
in the four cases, showing that the CMB temperature measurement is
sufficient to constrain $\xi$.}
\label{deltaxi}
\end{figure}

\begin{figure}
\caption{$\delta m / m$ versus $\xi$, for one family of massive
degenerate neutrinos.  The upper curves refer to $m= 0.07$ eV, the
lower curves for $m= 1$ eV. In each case, we give the results, from
top to bottom, for Planck without and with polarization, and Planck +
SDSS with $k_{max}=0.1$ or $0.2 h$ Mpc$^{-1}$.}
\label{deltamu}
\end{figure}
\newpage
\thispagestyle{empty}
\centerline{\epsfxsize=12cm\epsfbox{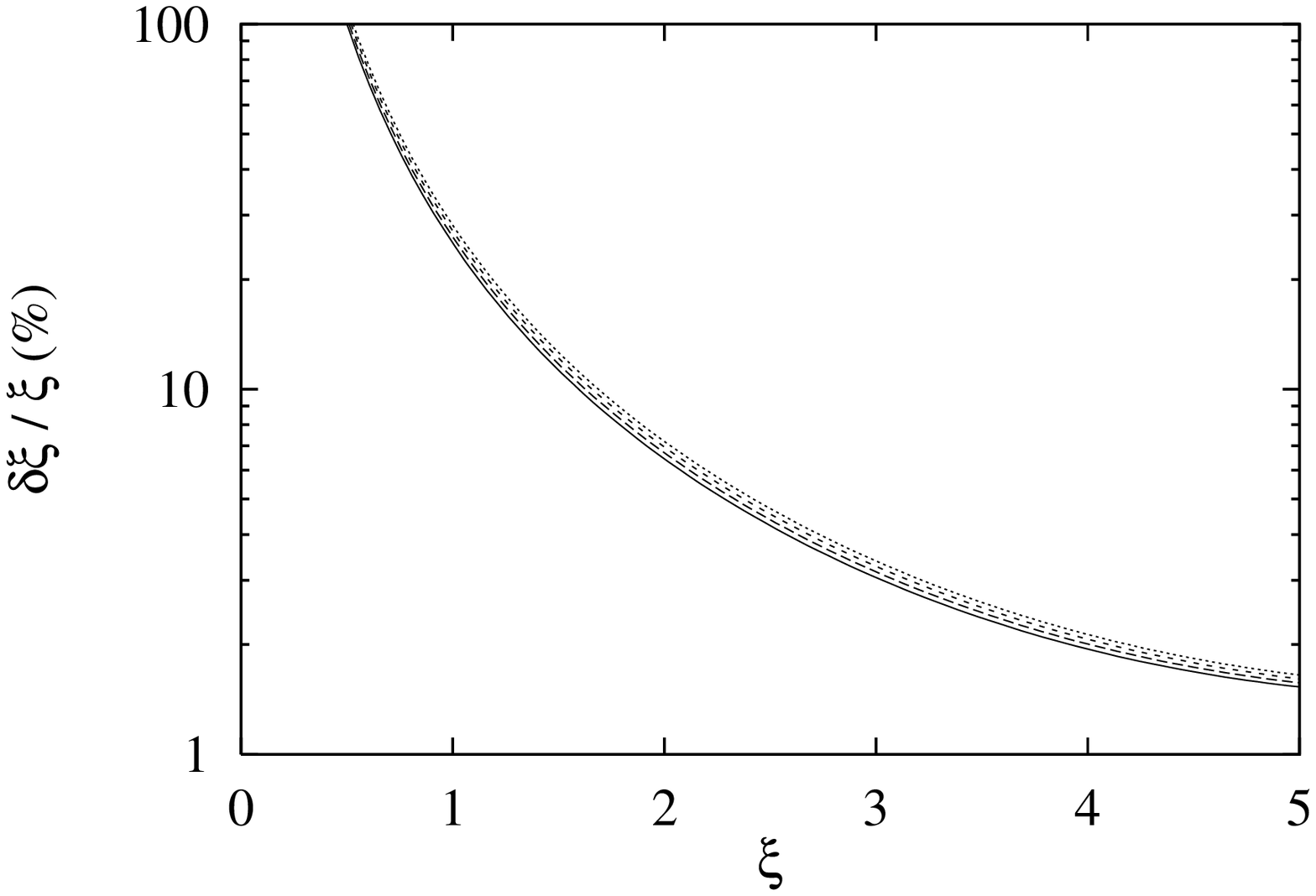}}
\vspace{2cm}
\centerline{Figure 1}
\newpage
\thispagestyle{empty}
\centerline{\epsfxsize=12cm \epsfbox{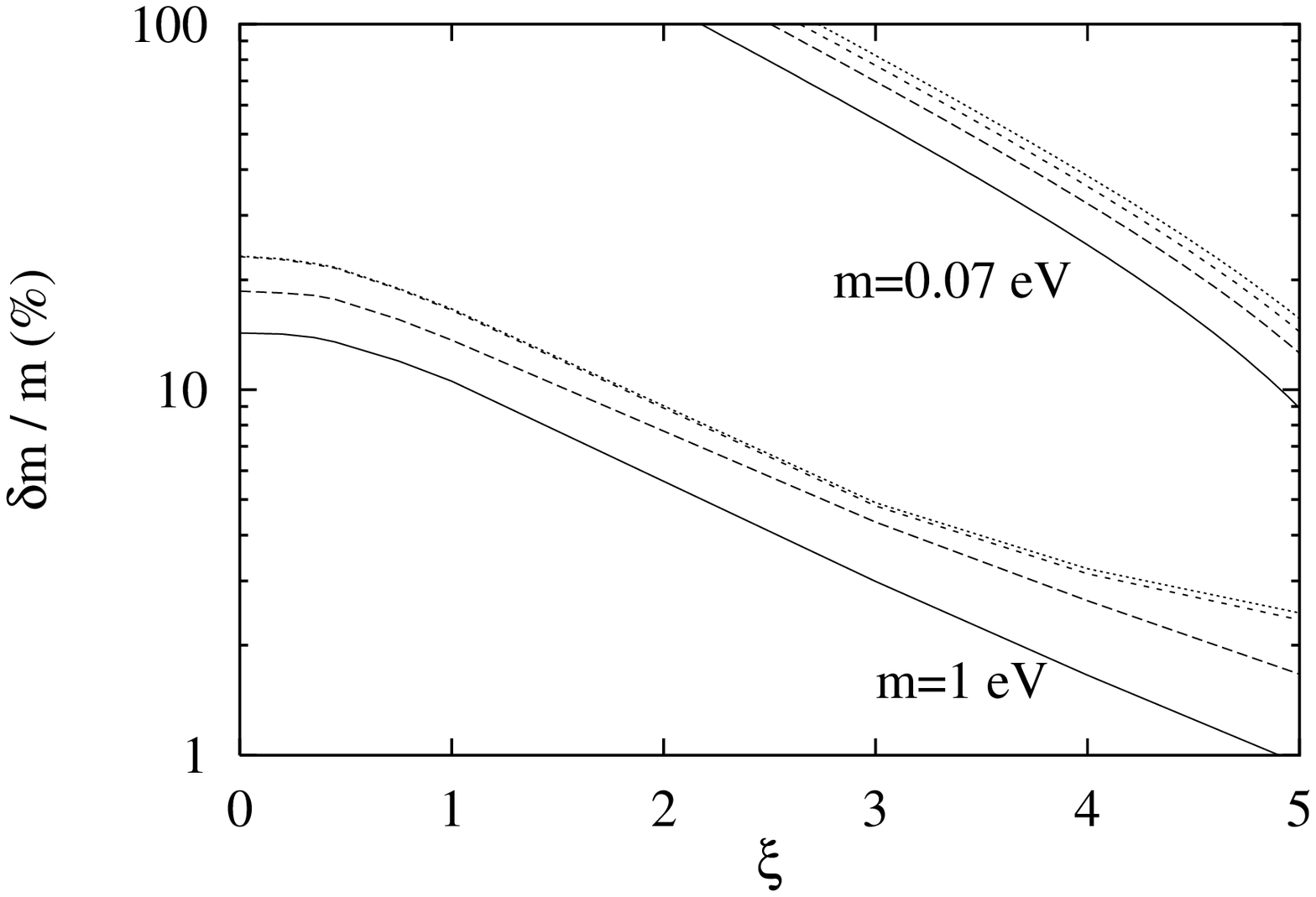}}
\vspace{2cm}
\centerline{Figure 2}%
\end{document}